\documentclass[10pt,letterpaper]{article}
\usepackage[top=0.85in,left=2.75in,footskip=0.75in]{geometry}

\usepackage{changepage}

\usepackage[utf8]{inputenc}

\usepackage{textcomp,marvosym}

\usepackage{fixltx2e}

\usepackage{amsmath,amssymb}

\usepackage{cite}

\usepackage{nameref,hyperref}

\usepackage[right]{lineno}

\usepackage{microtype}
\DisableLigatures[f]{encoding = *, family = * }

\usepackage{rotating}

\raggedright
\usepackage[aboveskip=1pt,labelfont=bf,labelsep=period,justification=raggedright,singlelinecheck=off]{caption}

\makeatletter
\renewcommand{\@biblabel}[1]{\quad#1.}
\makeatother

\date{}

\usepackage{lastpage,fancyhdr,graphicx}
\usepackage{epstopdf}
\fancyheadoffset[L]{2.25in}
\fancyfootoffset[L]{2.25in}
\begin{document}
\vspace*{0.35in}

\begin{flushleft}
{\Large
\textbf\newline{PGR: A Graph Repository of Protein 3D-Structures}
}
\newline
\\
Wajdi Dhifli\textsuperscript{1},
Abdoulaye Banir\'e Diallo\textsuperscript{1,*}
\\
\bigskip
\bf{1} Computer Science department, University of Quebec at Montreal, Downtown station, Montreal, Quebec, Canada, H3C 3P8
\\
\bigskip

%
%





* diallo.abdoulaye@uqam.ca

\end{flushleft}
\section*{Abstract}
Graph theory and graph mining constitute rich fields of computational techniques to study the structures, topologies and properties of graphs. These techniques constitute a good asset in bioinformatics if there exist efficient methods for transforming biological data into graphs. In this paper, we present Protein Graph Repository (PGR), a novel database of protein 3D-structures transformed into graphs allowing the use of the large repertoire of graph theory techniques in protein mining. This repository contains graph representations of all currently known protein 3D-structures described in the Protein Data Bank (PDB). PGR also provides an efficient online converter of protein 3D-structures into graphs, biological and graph-based description, pre-computed protein graph attributes and statistics, visualization of each protein graph, as well as graph-based protein similarity search tool. Such repository presents an enrichment of existing online databases that will help bridging the gap between graph mining and protein structure analysis. PGR data and features are unique and not included in any other protein database. The repository is available at \href{http://wjdi.bioinfo.uqam.ca/}{http://wjdi.bioinfo.uqam.ca/}.



\section*{Introduction}
\begin{figure}[!htpd]
\centering
\includegraphics[width=1\textwidth]{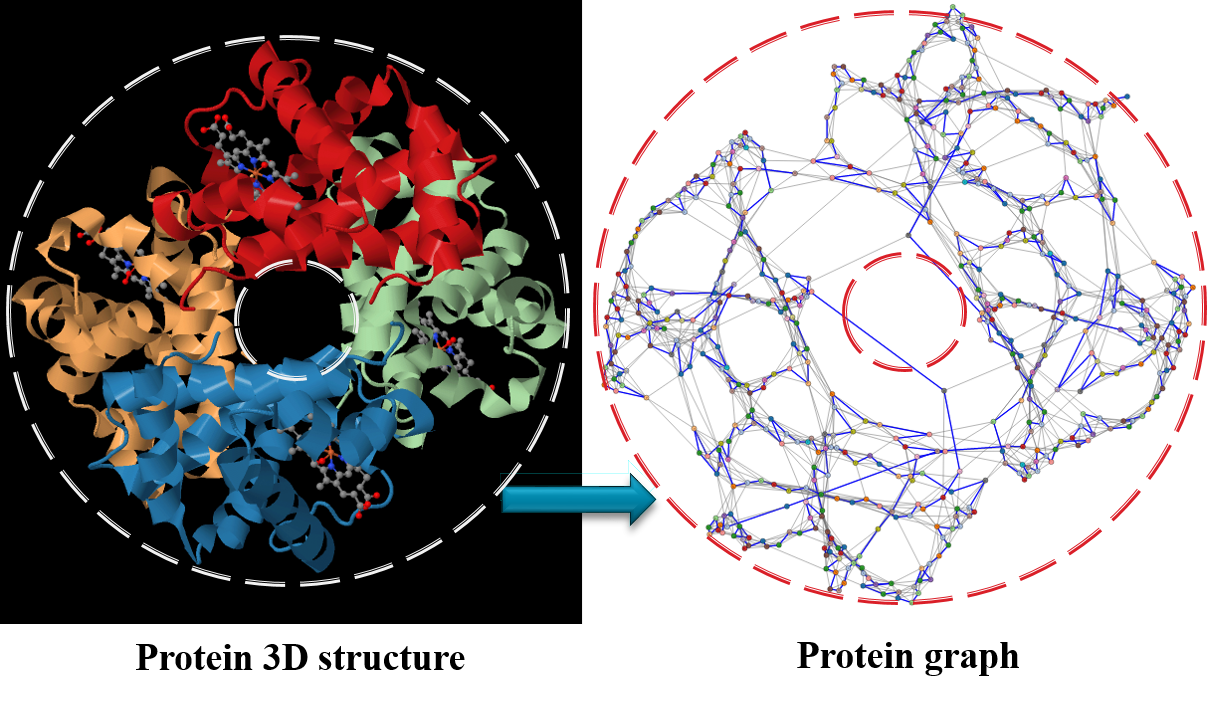}
\caption{{\bf The Human Hemoglobin protein 3D structure (PDB-ID: 1GZX) and its corresponding graph.} Each node in the graph represents an amino acid and each edge represents a spatial link (interaction) between two amino acids in the structure. The blue edges represent links from the primary structure and gray edges represent spatial links between amino acids that are distant in the primary structure. The Human Hemoglobin protein is composed of four subunits at the corners across a cavity at the center of the molecule. The example shows that the graph representation preserves the overall structure of the protein.} \label{fig:protein_to_graph}
\end{figure}

The advances in computational and biological techniques of protein studies have yielded enormous online databases. However, the complexity of protein structure requires adequate bioinformatics methods to mine these databases. The principles of graph theory have been adopted to investigate organic molecules \cite{Borgelt_2002} and proteins \cite{Milik_2003, Huan_2005, Dhifli_2014}. The tertiary structure captures homology between proteins that are distantly related in evolution. With the availability of more protein 3D-structures due to techniques such as X-ray crystallography, increasing efforts have been devoted to directly deal with them. 
%
%
%
A crucial step in the computational study of protein structures is to look for a convenient representation of their spatial conformations. The PDB format \cite{Berman_2000} represents the standard computer analyzable format that is used in online databases for representing macromolecular structures.  Extensions of the PDB format have been proposed in the literature mainly mmCIF and PDBML/XML file formats \cite{Westbrook_2005}. The PDB format and its extensions mainly consist on spatial coordinates of atoms composing the considered macromolecule besides its biological description and experimental details with which it was obtained. Such representation prevent a direct use of the large repertoire of available data mining and graph theory tools to study protein structures. A possible representation of protein 3D-structure can be a graph of interconnected amino acids. Figure \ref{fig:protein_to_graph} shows a real world example of the human hemoglobin protein and its corresponding graph. The graph representation preserves the overall structure and its components. Such representation can be considered as an alternative to existent representations, such as the PDB format \cite{Berman_2000}. 
%
\begin{figure}[!htpd]
\centering
\includegraphics[width=1\textwidth]{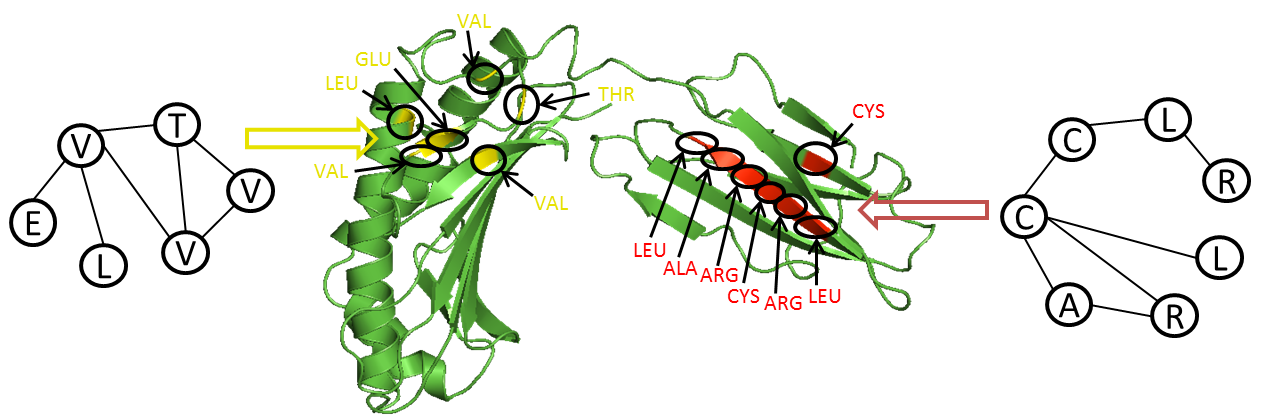}
\caption{{\bf An example of two subgraphs corresponding to two recurrent substructures extracted from a dataset of 38 proteins (including the HFE(human) hemochromatosis protein) from the immunoglobin C1-set domains family.} All the 38 proteins were transformed into graphs using PG-converter, then a frequent subgraph discovery was performed to discover recurrent substructures with a minimum support threshold of 30\%. This example shows the mapping of both subgraphs on the original 3D-structure of the HFE(human) hemochromatosis protein.} \label{fig:motif}
\end{figure}
It allows to fully exploit the potential of data mining and graph theory algorithms to perform complex studies such as the discovery of important substructures in protein families which can be performed through frequent subgraph mining, pattern recognition, and functional motif discovery. Figure \ref{fig:motif} shows a real example of two subgraphs corresponding to two recurrent substructures in a dataset of 38 proteins from the immunoglobin C1-set domains family, and their corresponding mapping on the original 3D-structure of the HFE(human) hemochromatosis protein. Such substructures are relevant for protein classification, protein function prediction, protein folding, $etc$. For instance, we have previously explored the potential of graph representation in the classification of four protein 3D-structure datasets, including G-proteins, immunoglobin C1-set domains, C-type lectin domains, and protein kinases catalytic subunit \cite{Dhifli_2014}. Frequent subgraphs were mined and used as features for the classification of each dataset. The experimental results showed that this graph-based approach outperformed the most competitive bioinformatics approaches including structural alignment-based classification (using Dali \cite{Holm_2010}) and Blast-based classification \cite{Altschul_1990}. 
%
%
In this paper, we present \textit{Protein Graph Repository} (PGR), an online repository of graphs representing all protein 3D-structures of the \textit{Protein Data Bank} (PDB)\cite{Berman_2000}. 
PGR provides bioinformatics tools that facilitate the integration of graph theory techniques in the core of protein 3D-structure studies \cite{Dhifli_2014, Stout_2008, Dognin_2014}. 

\section*{Materials and Methods}
\subsection*{Graph transformation of protein 3D structure}
Chemical interactions are the electrostatic forces that hold atoms and residues together, stabilizing proteins and forming molecules that give them their 3 dimensional shape \cite{Huan_2005, Stout_2008, Lovell_2003, Saidi_2009}. These interactions are mainly:
\begin{itemize}
\item[-] Covalent bonds between two atoms sharing a pair of valence electrons,
\item[-] Ionic bonds of electrostatic attractions between oppositely charged components,
\item[-] Hydrogen bonds between two partially negatively charged atoms sharing a partially positively charged hydrogen,
\item[-] Hydrophobic interactions where hydrophobic amino acids in the protein closely associate their side chains together,
\item[-] Van der Waals forces which represent transient and weak electrical attraction of one atom for another when electrons are fluctuating.
\end{itemize}

These interactions are supposed to be, in one form or another, the chemical analogues of the graph edges. Existing transformation approaches of protein 3D-structure into graph, similarly consider amino acids as graph nodes, but they differ in considering the edges in attempt to reflect the truly existing interactions. 
In the following, we present the main approaches in the literature that are used for building protein graphs in PGR. 
Let $G$ be a graph, $u$ and $v$ two nodes of $G$ ($u$,$v$ $\in$ $G$), $\Delta$ a function that computes the distance between pairs of nodes $\Delta(u,v)$, and $\delta$ a distance threshold.
\begin{itemize}
\item[-]\textbf{Main Atom} abstracts each amino acid in only one main atom, $M_A$ \cite{Huan_2005, Lovell_2003}. Two nodes representing two amino acids $u$ and $v$ are linked by an edge $e(u, v)$ if the euclidean distance between their two main atoms $\Delta(M_A(u), M_A(v))$ is below a distance threshold $\delta$. The main atom used in the literature is \textit{$C_\alpha$} with usually $\delta \geq 7 $\AA\textit{ } on the argument that $C_\alpha$ atoms define the overall shape of the protein conformation \cite{Huan_2005}.
\item[-]\textbf{All Atoms} considers the distances between all pairs of atoms $\Delta(A_A(u), A_A(v))$, where $A_A(u)$ represents all atoms of $u$ ($A_A(u)=\forall atom \in u$)\cite{Saidi_2009}. Two nodes representing two amino acids $u$ and $v$ are linked by an edge $e(u, v)$ if the euclidean distance between any pair of atoms from both amino acids $\Delta(A_A(u), A_A(v))$ is below a distance threshold $\delta$. Although this increases the complexity of graph building, it allows detecting connections that were omitted using Main Atom \cite{Saidi_2009}.
\end{itemize}

%
%

\subsection*{Main features of PGR}
\begin{figure}[!htpd]
\centering
\includegraphics[width=1\textwidth]{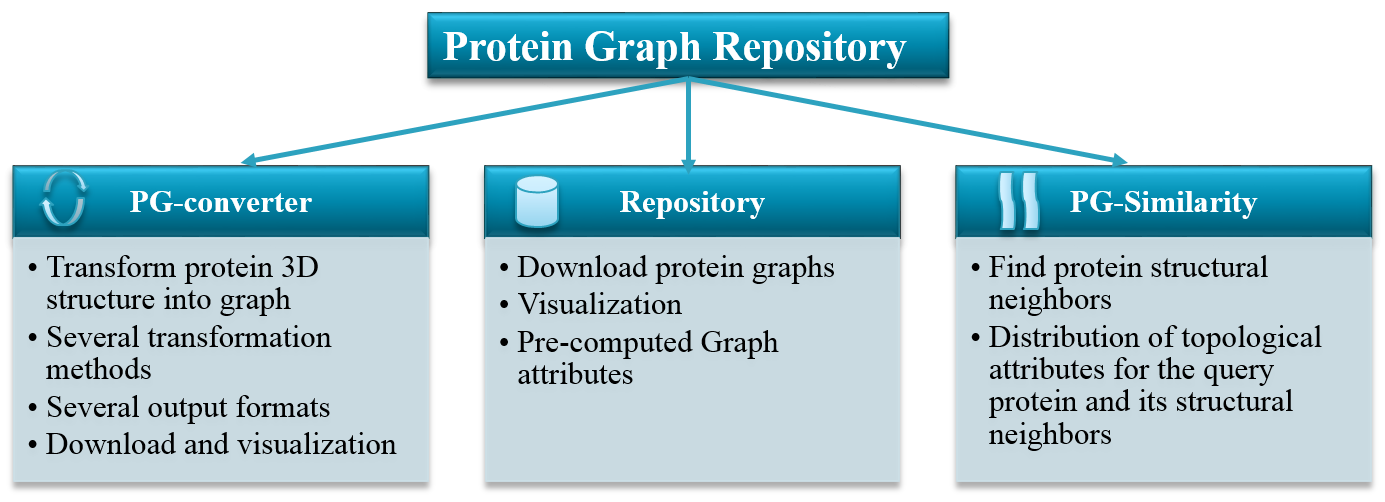}
\caption{{\bf PGR main bioinformatics features.}} \label{fig:pgr_features}
\end{figure}

The main bioinformatics features of PGR are listed in Figure \ref{fig:pgr_features}.

\paragraph{Graph repository} 
We transformed all protein 3D-structures (of the PDB) into protein graphs (in PGR) using the described methods. The repository is enriched by a selection tool allowing the filtering and targeting of a specific population of proteins. 
Each protein graph can be displayed solely in a light-weight and interactive visualization interface using the best available visualization libraries including D3.js \cite{Bostock_2011} and Cytoscape \cite{Smoot_2011}. A set of the most important attributes for protein graph mining have been pre-computed including density, diameter, link impurity, $etc$. These attributes are presented with their Z-score according to all protein graph attribute distributions. 
\paragraph{PG-converter} 
PGR also provides an online converter that allows to upload and transform protein 3D-structures into protein-graphs. Available transformation methods include All Atoms, Main Atom based on C$\alpha$, C$\beta$, amino acid centroid, side chain centroid, amino acid ray, amino acid ray and side chain orientation, or side chain ray. 
\paragraph{PG-similarity}
Furthermore, we provide a search tool based on the pairwise similarity of structural protein attributes. Such tool could constitute an asset for several biological tasks such as protein classification and function prediction. The pairwise similarity between two protein graphs is measured by the distance between their corresponding vector representation based on the structural and topological attributes. We selected a set of attributes from the literature that are interesting and efficient in describing connected graphs \cite{Li_2012, Leskovec_2005}. In the following, we list and define the used structural and topological attributes:
\begin{enumerate}
\item \textbf{Number of nodes}: The total number of nodes in the protein graph, also called the graph order $|V|$.
\item \textbf{Number of edges}: The total number of edges in the protein graph, also called the graph size $|E|$.
\item \textbf{Average degree}: The degree of a node $u$, denoted $deg(u)$, represents the number of nodes adjacent to $u$. The average degree of a graph $G$ is the average value of the degrees of all nodes in $G$. Formally: $ deg(G) = \frac{1}{n} \sum^n_{i=1} deg(u_i)$ where $deg(u_i )$ is the degree of the node $u_i$ and $n$ is the number of nodes in $G$. 
\item \textbf{Density}: The density of a graph $G=(V, E)$ measures how many edges are in $E$ compared to the maximum possible number of edges between the nodes in $V$. Formally: $ den(G) = \frac{2 \mid E\mid}{(\mid V\mid\ast (\mid V\mid -1))}$.
\item \textbf{Average clustering coefficient}: The clustering coefficient of a node $u$, denoted by $c(u)$, measures how complete the neighborhood of $u$ is, $i.e.$, $c(u)= \frac{2 e_u}{k_u (k_u - 1)}$ where $k_u$ is the number of neighbors of $u$ and $e_u$ is the number of connected pairs of neighbors. If all the neighbor nodes of u are connected, then the neighborhood of $u$ is complete and we have a clustering coefficient of 1. If no nodes in the neighborhood of $u$ are connected, then the clustering coefficient is 0. The average clustering coefficient of an entire graph $G$ having $n$ nodes, is given as the average value over all the nodes in $G$. Formally: $C(G)= \dfrac{1}{n} \sum_{i=1}^n c(u_i)$.
\item \textbf{Average effective eccentricity}: For a node $u$, the effective eccentricity represents the maximum length of the shortest paths between $u$ and every other node $v$ in $G$, $i.e.$, $e(u) = max\{d(u,v) : v\in V\}$. If $u$ is isolated then $e(u) = 0$. The average effective eccentricity is defined as $Ae(G)= \frac{1}{n}\sum_{i=1}^n e(u_i)$, where $n$ is the number of nodes of $G$.
\item \textbf{Effective diameter}: The effective diameter represents the maximum value of effective eccentricity over all nodes in the graph $G$, $i.e.$, $diam(G) = max\lbrace e(u)\mid u\in V\rbrace$ where $e(u)$ represents the effective eccentricity of $u$ as defined above.
\item \textbf{Effective radius}: The effective radius represents the minimum value of effective eccentricity over all nodes in the graph $G$, $i.e.$, $rad(G) = min\lbrace e(u)\mid u\in V\rbrace$ where $e(u)$ represents the effective eccentricity of $u$.
\item \textbf{Closeness centrality}: The closeness centrality measures how fast information spreads from a given node to other reachable nodes in the graph. For a node $u$, it represents the reciprocal of the average shortest path length between $u$ and every other reachable node in the graph, $i.e.$, $C_c(u) = \frac{n-1}{\sum_{v\in \lbrace V\setminus u\rbrace} d(u,v)}$ where $d(u,v)$ is the length of the shortest path between the nodes $u$ and $v$. For a graph $G$, we consider the average value of closeness centrality of all the nodes, $i.e.$, $C_c(G) = \frac{1}{n} \sum_{i=1}^n u_i$.
\item \textbf{Percentage of central nodes}: Here, we compute the ratio of the number of central nodes from the number of nodes in the graph. A node $u$ is considered as central point if the value of its eccentricity is equal to the effective radius of the graph, $i.e.$, $e(u) = rad(G)$.
\item \textbf{Percentage of end points}: It represents the ratio of the number of end points from the total number of nodes of the graph. A node $u$ is considered as end point if $deg(u) = 1$.
\item \textbf{Neighborhood impurity}: The impurity degree of a node $u$ belonging to a graph $G$, having a label $L(u)$ and a neighborhood (adjacent nodes) $N(u)$, is defined as $ImpurityDeg(u) = \mid L(v): v \in N(u), L(u)\neq L(v)\mid$. The neighborhood impurity of a graph $G$ represents the average impurity degree over all nodes with positive impurity.
\item \textbf{Link impurity}: An edge $\{u,v\}$ is considered to be impure if $L(u)\neq L(v)$. The link impurity of a graph $G$ with $k$ edges is defined as: $\frac{\mid\{u,v\}\in E: L(u)\neq L(v)\mid}{k}$.
\item \textbf{Label entropy}: It measures the uncertainty of labels. The label entropy of a graph $G$ having $k$ labels is measured as $E(G) = -\sum_{i=1 }^k p(l_i)\textit{ log }p(l_i)$, where $l_i$ is the $i^{th }$ label.

\end{enumerate}


\subsection*{Illustrative Example}
\begin{figure}[!htpd]
\centering
\includegraphics[width=1\textwidth]{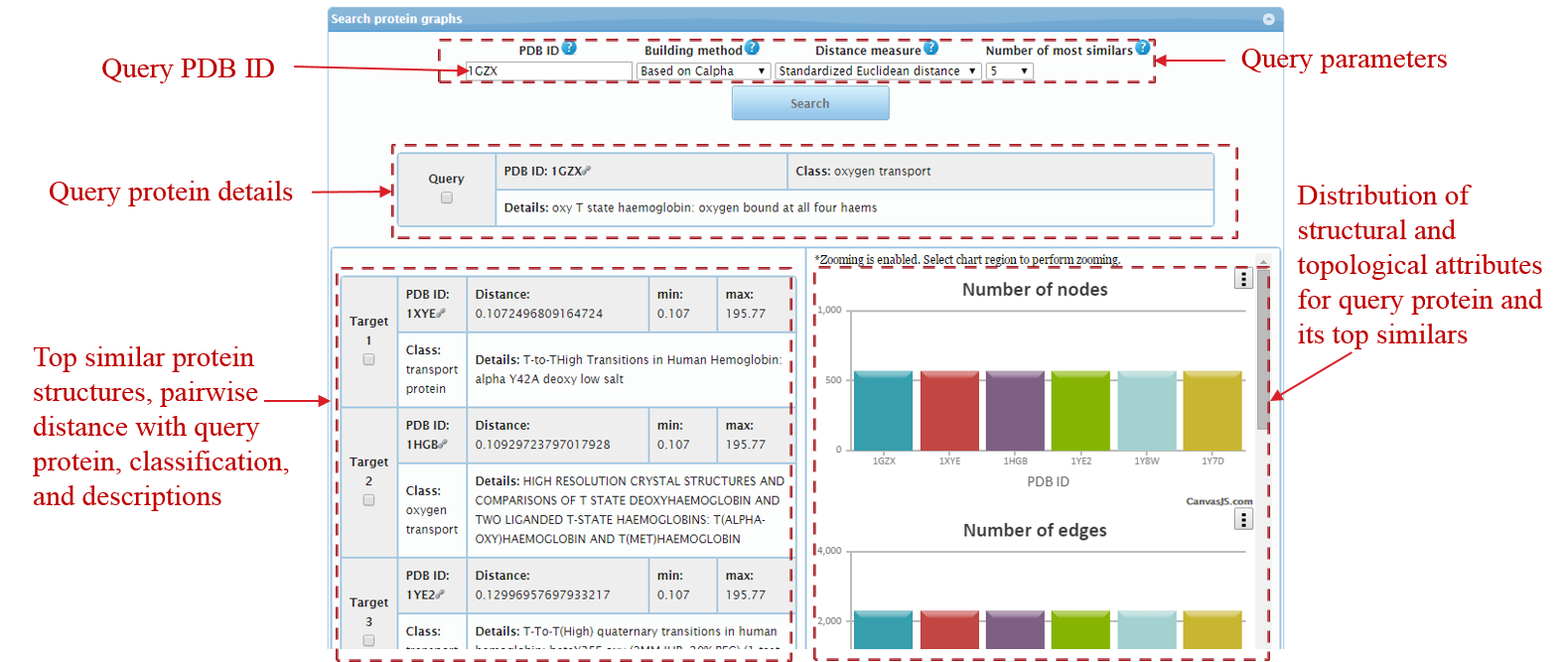}
\caption{{\bf Example of PG-Similarity search of structural neighbors of the Human Hemoglobin protein (PDB-ID: 1GZX).}} \label{fig:pg_similarity_example}
\end{figure}

This section shows an illustrative example of a graph similarity search for the \textit{Human Hemoglobin protein} (Figure \ref{fig:protein_to_graph}) using PG-similarity. Figure \ref{fig:pg_similarity_example} shows the different components of PG-similarity web interface. The graph similarity search based on the set of structural and topological attributes is performed to detect the top similar proteins for the considered tertiary structure. The graph representation used for the query is based on C$\alpha$ and the vector distance measure is the standardized euclidean distance. 

Table \ref{tab:pg_similarity_example} shows the obtained results for the top 10 most similar proteins. Similarly to the query 3D structure, all the obtained proteins are also Hemoglobin molecules and they are part of the same organism of the query protein, namely Homo sapiens. The similarity search was only based on the previously described structural and topological attributes with no additional knowledge about nor the query, neither the target structures. This demonstrates that PG-similarity allows an accurate detection of top similar proteins that is biologically meaningful. The selected similar could constitute an asset for several biological tasks such as protein classification and function prediction.

\begin{table}[!htpd]
\caption{
{\bf Results of PG-similarity search for the top 10 similar structures to the Human Hemoglobin protein (PDB ID: 1GZX).}}
\begin{tabular}{|l|l|l|l|l|l|l|l|}
\hline
{\bf Rank} & {\bf PDB ID} & {\bf Distance} & {\bf Classification} & {\bf Molecule} & {\bf Taxonomy} \\ \hline
1 & 1XYE & 0.1072 & Transport protein & Hemoglobin & Homo sapiens \\ \hline
2 & 1HGB & 0.1092 & Oxygen transport & Hemoglobin & Homo sapiens  \\ \hline
3 & 1YE2 & 0.1299 & Transport protein & Hemoglobin & Homo sapiens   \\ \hline
4 & 1Y8W & 0.1319 & Transport protein & Hemoglobin & Homo sapiens  \\ \hline
5 & 1Y7D & 0.1353 & Transport protein & Hemoglobin & Homo sapiens  \\ \hline
6 & 1Y7C & 0.1357 & Transport protein & Hemoglobin & Homo sapiens  \\ \hline
7 & 1DXV & 0.1395 & Oxygen transport & Hemoglobin & Homo sapiens  \\ \hline
8 & 1XZ7 & 0.1475 & Transport protein & Hemoglobin & Homo sapiens  \\ \hline
9 & 1YGD & 0.1493 & Transport protein & Hemoglobin & Homo sapiens  \\ \hline
10 & 1XXT & 0.152 & Transport protein & Hemoglobin & Homo sapiens  \\ \hline
\end{tabular}
\begin{flushleft}
The used parameters for the query are: query PDB ID: 1GZX, graph building method: based on C$\alpha$, distance measure: standardized euclidean distance, number of most similars: 10. 
\end{flushleft}
\label{tab:pg_similarity_example}
\end{table}

\section*{Results and Discussion}
Regarding the graph transformation of protein 3D structures, both Main Atom and All Atoms suffer drawbacks. Since, the Main Atom technique abstracts amino acids into one main atom, it may omit possible edges between other atoms in the amino acids that are more close than their main atoms. Moreover, in the case of considering centroids of the amino acids as the main atoms, it may also suffer from two problems. In the case of two big amino acids, if their centroids are farther than the given distance threshold, they will be considered with no links while a real connection could be established between other close atoms (other than the centroids). In the case of small amino acids, if the distance between their centroids is smaller than the given distance threshold then they will be considered as connected while they can be disconnected in reality. The all Atoms technique overcomes both limitations by theoretically considering the distance between all the atoms in the amino acids, this highly increases the runtime and complexity of the technique. However, the authors proposed some heuristics to alleviate the complexity. For instance, they consider only the distance between the side chains' centroids to decide whether their amino acids are connected or not, without regards to their chemical properties. This reduces the runtime but it may engender false edges.

Compared to the conventional distance matrix representation \cite{Holm_2010} \textit{PG-similarity} measures the global structural and topological similarity between protein structures on a macro side, whereas distance matrix based similarity operates on a micro side and looks into every single detail in compared structures. Even though both similarity methods should be highly correlated and not diverge (as similar structures have similar topological descriptions), each method has its positive and negative sides. Distance matrix based methods has the advantage of detecting exact superposition and local matching sites, however, they are combinatorial and thus computationally costly. PG-similarity method is based on a vector embedding of graphs of protein structures based on a set of structural and topological attributes. This makes it unable to return local matches, however, such strategy makes it able to capture structural similarity in a very fast way. Moreover, some attributes, like clustering coefficient and neighborhood impurity, makes PG-similarity able to reveal hidden similarities that are undetected using existing methods.

With the growth of protein 3D-structures in online databases, the transformation of protein 3D-structures into graphs of interconnected amino acids and the application of graph mining concepts constitute a relevant feature for the development of rapid and efficient computational techniques. So far, PGR contains 188 252 graphs corresponding to 94 126 protein 3D-structures from the PDB. The 188 252 protein graphs are composed of 94 126 graphs created using Main Atom method with $C_\alpha$ and 94 126 graphs created using All Atoms method. PGR data will be regularly updated according to the PDB. PGR is an independent repository, but it can also act as a complementary resource to existing ones such as the PDB. PGR is apt for extension and additional services and functionalities will be added in the next coming versions.

%
%
%
\section*{Acknowledgments}
This work is supported by a grant of the Fonds de recherche du Qu\'ebec - Nature et technologies to A.B. Diallo.


%
%
%

\end{document}